\documentclass[12pt]{iopart}

\begin{document}

\title[Discrete Wigner distribution for two qubits:
a characterization...]{Discrete Wigner distribution for two
qubits: a characterization of entanglement properties}

\author{Riccardo Franco
\footnote[3]{To whom correspondence should be addressed
riccardo.franco@polito.it}, Vittorio Penna}
\address{Dipartimento di Fisica and U.d.R. I.N.F.M., Politecnico di Torino
C.so Duca degli Abruzzi 24, I-10129 Torino, Italia}

\date{\today}

\begin{abstract}
We study the properties of the discrete Wigner distribution for
two qubits introduced by Wotters. 
In particular, we analyze the entanglement properties within the
Wigner distribution picture by considering the negativity of the
Wigner function (WF) and the correlations of the marginal distribution.
%
%
We show that a state is entangled if at least one among the values
assumed by the corresponding discrete WF is smaller than a certain
critical (negative) value.
Then, based on the Partial Transposition criterion, we establish the
relation between the separability of a density matrix and the
non-negativity of the WF's relevant both to such a density matrix and 
to the partially transposed thereof.
%
%
Finally, we derive a simple
inequality --involving the covariance-matrix elements
of a given WF-- which appears to provide
a separability criterion stronger than the
one based on the Local Uncertainty Relations.

\end{abstract}

\maketitle

%
\section{Introduction}
A quantum system with continuous degrees of freedom can be
represented in terms of a Wigner Function \cite{Wigner} defined as
a real function on the phase space. The Wigner function (WF) is
similar to a probability distribution (its integration over the
phase space is normalized to one) even if it can take negative
values on restricted domains. There is an extensive literature on
the continuous-system WF \cite{HOSW}-\cite{NarConn}
due to its wide applicability in different contexts of physics.
Concerning the WF of discrete systems, the literature is less
extensive even if this theme has recently attracted a lot of
interest mainly in view of the role that a discrete phase-space
structure can play within Quantum Information Theory
\cite{Wootters}, \cite{Wootters2003}-\cite{KBJ}. In this respect
paper \cite{Wootters2004} contains a useful list of references.
Different generalizations of the WF to quantum systems with a
finite-dimensional Hilbert space have been proposed in the
literature such as 1) the continuous WF for spin variables
\cite{Stratonovich} and 2) the definitions of WF based on a
discrete phase space. As to the latter, early studies
were made in \cite{CohenScully, Fey}. 
A discrete WF has been introduced in \cite{Wootters} and \cite{Galetti-DeTol}
which generalizes the $2\times 2$ case of \cite{Fey}
and is valid for systems having a $N$-dimensional Hilbert space, 
with $N$ a prime or a power of a prime. More recently,
an alternative definition of WF involving Galois fields 
\cite{Wootters2003}-\cite{Vourdas2005}
has allowed the study of the composite-dimensional case
and evidenced several interesting tomographic properties
\cite{Wootters2003,Colin}.

The present article is focused on studying the entanglement
properties of a two-qubit system by using the
discrete WF defined in \cite{Wootters}. It is worth noting that
two-qubit (and more in general many-qubit) systems 
have received an increasing attention, not only within spin models,
but also in the recent literature
on optically trapped bosons modelled within the
Bose-Hubbard picture \cite{spinbos}. 
The impressive experimental progress in
controlling the spatial trapping of bosons makes the realization
of many-qubit systems a quite realistic objective.
In this paper we find some new separability criteria and recast
other known criteria in terms of discrete WF's. We check them 
evaluating two-qubit entanglement and show that the discrete WF 
describes both classical and quantum correlations
better than the density-matrix approach. 
We note how using the WF not only improves the
visualization of the system state but is also expedient experimentally: 
since the WF is directly related to tomographic techniques, the separability
criteria coming from the WF do not require to know of all the matrix elements.

In section 2, we review the definition of discrete phase-space
given in reference \cite{Wootters} --this is particularly useful
for our purposes-- and the corresponding discrete Wigner and
Characteristic functions. We present some basic properties of
these functions which provide a useful tool for studying quantum
correlations. In section 3 we consider four different separability
criteria in terms of the two-qubit discrete WF and of its
covariance matrix. In particular, we show that 1) there is a
negative value of discrete WF that allows one to discriminate
between separable and entangled states, 2) the Partial
Transposition criterion can be reformulated in terms of the
two-qubit discrete WF, 3) there is a nontrivial link between the
separability of the density matrix and the non-negativity of the
WF's corresponding both to the density matrix and to its partial
transposed matrix, 4) the Local Uncertainty Relations
relevant to phase space operators can be generalized in terms of
the WF covariance matrix (thus evidencing the difference between
classical and quantum correlations), and 4) the Generalized
Uncertainty Principle, so far studied for WF's relevant to
continuous phase space, is extended to the case of a discrete WF.
%
%
\section{Discrete phase space and Wigner function}\label{discrete_phase_space}
In a discrete $r$-dimensional Hilbert space, with $r$ a prime
number, the phase space can be defined \cite{Wootters} as a
$r\times r$ array of points. The latter can be labelled by pairs
of coordinates $\alpha=(q,p)$, each taking values from 0 to $r-1$.
For each coordinate we define the usual addition and
multiplication mod $r$ thus obtaining the structure of a finite
mathematical field ${\bf F}_{r}$ with $r$ elements
$(0,1,...,r-1)$. If the dimension is $N=r^n$, with $r$ prime and
$n$ an integer greater than 1, the discrete phase space can be
built in two ways, both giving a discrete phase space formed by a $N
\times N$ grid: the first involves the extension ${\bf F}_N$ of
the primitive field ${\bf F}_r$ \cite{Wootters2004,
Wootters2004_bis}, while the second is based on performing the
$n$-fold cartesian product of $r\times r$ phase spaces
\cite{Wootters}.
In the present article we will use this last definition of
discrete phase-space (entailing the definition of WF given in
\cite{Wootters}). The choice of phase-space structure is justified
by the direct connection with the tensor-product structure of the
Hilbert space ensuing from the decomposition of the system in two
or more subsystems, which is a useful feature for studying the
entanglement. According to this definition, the phase-space points
$\alpha$ are labelled as $n$-tuple $(\alpha_1, \alpha_2, ... ,
\alpha_n)$ of coordinates, each $\alpha_i$ pertaining to the
$i$-th subsystem. In each subsystem with prime dimension $r$ we
can build standard lines as set of points satisfying equation
$(uq+vp)_{mod r}=c$. However, we cannot define uniquely lines over
the entire phase space (with modular arithmetic): in reference
\cite{Wootters2004} there is an example of two sets of points
which form two parallel "lines" but intersect in two distinct
points. Nevertheless, we can define the alternative concept of
\emph{slice} \cite{Wootters}: given a set of $n$ lines
$\{\lambda_i\}$ (one for each subsystem), the slice is the set of
all points $\alpha = (\alpha_1, \alpha_2, ..., \alpha_n)$, where
$\alpha_i \in \lambda_i$. A weaker notion of parallelism can be
defined: two slices are parallel if each of the $n$ lines forming
the first slice are parallel to the corresponding $n$ lines of the
second slice.
%
%
\subsection{Definition of the discrete WF}\label{WFdefinition}
The discrete WF relevant to a $N=r^n$-dimensional system can be
defined \cite{Wootters} by means of the set of discrete
phase-point operators $\widehat{A}({\alpha})$ (or $\Delta(\alpha)$
\cite{Klim-mun} ).
Consistently with the definition of phase space in terms 
(of cartesian product) of constituent subspaces  \cite{Wootters},
phase-point operators $\widehat{A}({\alpha})$ are defined as
tensor product of phase-point operators relevant to the
corresponding subsystems:  $\widehat{A}({\alpha}) \, =
\widehat{A}({\alpha_1}) \otimes \widehat{A}({\alpha_2}) \otimes
... \otimes \widehat{A}({\alpha_n})$. Since they form a complete
orthogonal basis for the Hermitian $N\times N$ matrices, any
density matrix can be written as $\widehat{\rho}=\sum_{\alpha}
W(\alpha)\widehat{A}(\alpha)$, where the real-valued coefficients
\begin{equation}\label{wign_discr_1}
W(\alpha)=\frac{1}{N}tr[\widehat{\rho} \widehat{A}(\alpha)]
\end{equation}
represent the discrete Wigner function (also called the discrete
Weyl symbol).
%
Phase-point operators exhibit two basic properties: i) for any
couple of points $(\alpha_1,\alpha_2)$
\begin{equation}\label{orth_A_alpha}
 tr[\widehat{A}(\alpha_1)\widehat{A}(\alpha_2)]=N\delta(\alpha_1, \alpha_2)\, ,
\end{equation}
%
ii) given any slice $\lambda$ in the phase space, the projector
relevant to $\lambda$ can be written as
\begin{equation}\label{Proj_A_alpha}
  \widehat{P}_{\lambda}=\frac{1}{N}\sum_{\alpha \in
       \lambda}\widehat{A}(\alpha)\, .
\end{equation}
The latter definition implies that the set of all
$\widehat{P}_{\lambda'}$ for which $\lambda'$ is parallel to
$\lambda$ forms a set of mutually orthogonal projection operators.
Moreover, given the slice $\lambda=(\lambda_1, \lambda_2, ...,
\lambda_n)$, then $\widehat{P}_{\lambda}$ is the tensor product
$\widehat{P}_{\lambda_1} \otimes \widehat{P}_{\lambda_2} \otimes
.. \otimes \widehat{P}_{\lambda_n}$ of projectors relevant to the
subsystems. Such properties, analogous to those characterizing
continuous phase-point operators \cite{Wootters}, can be used to
derive the discrete-WF properties.
Owing to formulas \ref{orth_A_alpha} and \ref{Proj_A_alpha}
discrete WF's feature two crucial properties. First, if
$W(\alpha)$, $W'(\alpha)$ correspond to density matrices $\rho$,
$\rho'$, respectively, then formula \ref{orth_A_alpha} entails
that
\begin{equation}\label{orth_wign_discr}
N\sum_{\alpha}W(\alpha)W'(\alpha)=tr(\rho\rho')\, .
\end{equation}
Second, due to equation \ref{Proj_A_alpha}, given a complete set
of $N$ parallel slices, for each slice $\lambda$, the $N$ real
numbers $p_{\lambda}=\sum_{\alpha \in \lambda}W_{\alpha}$ are the
probabilities of the outcomes of a specific measurement associated
with $\lambda$. Hence $\sum_{\alpha}W_{\alpha}=1$ (normalization
property).

Let us consider first the simple case 
$N=2$ (single qubit). The phase-point operators can be written
in terms of Pauli matrices as
\cite{Wootters}
\[ \fl 
\widehat{A}(\alpha)=\frac{1}{2}\left[I+(-1)^{q}
{\sigma}_z + (-1)^{p} {\sigma}_x+(-1)^{q+p}
{\sigma}_y\right] \, ,
%
%
\quad
\begin{tabular}{cc}
  $\sigma_x=\left[
    \begin{tabular}{cc}
    0 & 1 \\
    1 & 0
    \end{tabular}\right]$\, ,
       &
  $\sigma_y=\left[
    \begin{tabular}{cc}
    0 & -i \\
    i & 0
    \end{tabular}\right]$
%
\end{tabular}
\]
and $\sigma_z= [\sigma_x , \sigma_y]/2i$.
The single-qubit phase space is the set of points $\alpha=(q,p)$,
where $q, p\, = 0, 1$, exhibiting properties of $\bf F _2$. Since
the density matrix for a general one-qubit state can be written
(within the standard computational basis)
in terms of three independent real elements $\rho_{00},
Re(\rho_{01}), Im(\rho_{01})$, we have
$  
W(q,p)=\frac{1}{4}\{[1+(-1)^{q}]\rho_{00}+[1-(-1)^{q}]\rho_{11}+(-1)^{p}2 Re(\rho_{01})+(-1)^{q+p}2 Im(\rho_{01})\}
$.
In the case of two qubits, the WF has a more complex expression.
Upon noting that the phase space operators are defined as
$\widehat{A}(\alpha_1,\alpha_2)=\widehat{A}({\alpha_1})\otimes
\widehat{A}({\alpha_2})$, the WF becomes
\begin{equation}\label{wign_discr_2}
 W(q_1, q_2, p_1, p_2)= \frac{1}{4}tr[\widehat{\rho} \widehat{A}(q_1, p_1)\otimes \widehat{A}(q_2, p_2)]\, .
\end{equation}
When necessary, we shall write $W_{\rho}$, where the subscript
means that the WF is associated to density matrix $\rho$.
%
%
\subsection{The discrete characteristic function}\label{Char_function_def}
The set $\{I,\sigma_x,\sigma_y,\sigma_z\}$ forms an orthogonal
basis for the set of hermitian operators acting on a single qubit.
Thus any density matrix $\rho$ for a single qubit can be written
as $\rho = \frac{1}{2}\sum_{uv}\chi(u,v)\widehat{S}(u,v)$ while
the characteristic function for a single qubit is
\begin{equation}
\label{char_discr_1}
 \chi(u,v)=tr[\rho \widehat{S}(u,v)]
\end{equation}
where $\widehat{S}(0,0)=I$, $\widehat{S}(1,0)=\sigma_x$,
$\widehat{S}(0,1)=\sigma_z$, and $\widehat{S}(1,1)=\sigma_y$. When
necessary, we write the argument $\beta$ instead of $(u,v)$, or
the single index $i$, where $i=u+2v$ assuming integer values from
$0$ to $3$. In the case of two qubits, any density matrix can be
written as $\rho =\frac{1}{4}\sum_{\beta_1 \beta_2} \chi
({\beta_1, \beta_2}) S({\beta_1}) \otimes S({\beta_2})$. The
two-qubit characteristic function is thus defined as $\chi
({\beta_1, \beta_2}) =tr[\rho \widehat{S}(\beta_1)\otimes
\widehat{S}(\beta_2)]$. Function $\chi(\beta)$ is connected with
the discrete WF by a discrete Fourier transform. For example, in
the single-qubit case ($r=2$),
$W(q,p)=\frac{1}{4}\sum_{\beta}(-1)^{(qu+pv)}\chi(u,v)$ , and in
the two-qubit case
\[ \fl
 W(q_1,q_2,p_1,p_2)=
 \frac{1}{16}\sum_{u_1,u_2,v_1,v_2}(-1)^{
 (q_1 u_1 + p_1 v_1)+(q_2 u_2 +p_2 v_2)}\chi(u_1,u_2,v_1,v_2) \, .
\]
It is worth noting that the determination of $\chi (\beta)$ is
connected to a specific tomographic technique \cite{Wootters2003}.
For example, in the case of spin $\frac{1}{2}$ particles, it
consists of repeated measures of spin "up vs down" along three
directions relevant to each particle. The WF can be determined via
the previous equations from the characteristic function.
The Inner Product Rule for $\chi (\beta)$ corresponding to formula
\ref{orth_wign_discr} is
\begin{equation}\label{orth_chi_discr}
  \frac{1}{N}\sum_{\beta}\chi(\beta)\chi'(\beta)=tr(\rho\rho')\, .
\end{equation}
%
%
Operators $\widehat{S}$ can be thought as translation operators.
The single-qubit WF relevant to
$\widehat{\rho'}=\widehat{S}(a,b)\widehat{\rho}\widehat{S}(a,b)^{\dag}$
is $W_{\rho'}(q,p)=W_{\rho}(q+a,p+b)$, where the sum in the
argument is mod2. In the two-qubit case, operators $\widehat{S}$
act as translation operators on each single-qubit phase space.
This formalism supplies a useful tool for recognizing the
translational covariance of the WF
\cite{Klim-mun,PazRoncagliaSaraceno}.

%
\subsection{Graphical representation, pure and mixed states.}\label{Prop-rep}
The phase space of the single qubit is represented in table
\ref{tb:Graph_Rep1}, left panel. A simple example of WF is given
by the spin $j={1}/{2}$ coherent state $|\xi \rangle =
\left(\left |0\right>+\xi\left|1\right>\right)/ \sqrt{1+|\xi|^2}$
(where $|0\rangle$ and $|1\rangle$ are the spin-up and the
spin-down state, respectively), which is known to be the most
general pure state for a single qubit. Given $|\xi \rangle$, the
corresponding WF is depicted in table \ref{tb:Graph_Rep1}, right
panel. The WF can have only one negative element (otherwise, at
least one probability associated to a direction would be negative)
determined by $|Re(\xi)+Im(\xi)|>1$ or
$|Re(\xi)-Im(\xi)|>|\xi|^2$.
An important observation is that,
while most positive value of
the single-qubit WF is $1/2$, the 
most negative value $(1-\sqrt{3})/4$ is assumed in correspondence 
to state $|\xi \rangle$ with $\xi=(1+i)/({1-\sqrt{3}})$ 
\cite{Wootters2003}. These extremal values are useful to
write a simple separability
criterion.

In table \ref{tb:Grap_Rep2}, we illustrate $W_\alpha
=W(q_1,q_2,p_1,p_2)$ on the discrete phase-space points for the
two qubit case, where the phase-space label is $\alpha=(\alpha_1,
\alpha_2)=(q_1,q_2,p_1,p_2)$. This table is useful to clarify
the notation we adopt for the WF (which differs from that of
reference \cite{Wootters2004}). 
%
The purity character of a state can be evinced both from the WF
and from the characteristic functions. From the general equations
\ref{orth_wign_discr} and \ref{orth_chi_discr}, we find
\begin{equation}
\label{W_chi_p-m}
  \frac{1}{N}\leq N\sum_{\alpha}W(\alpha)^2=
\frac{1}{N}\sum_{\beta}\chi(\beta)^2=
  tr(\widehat{\rho}^2)\leq1 \, ,
\end{equation}
where the equality holds for pure states whereas the inequality
is involved by mixed states. In order to define a "mixed state"
we recall that, upon introducing the basis 
$\{ | e \rangle : \rho | e \rangle = p_e | e \rangle  \}$
relevant to a given density matrix $\rho$, a state is "mixed"
when more than one eigenvalue $p_e$ is nonzero. In this case
the system state is represented by $\rho =\sum_e p_e | e \rangle\langle e|$ 
where probabilities $p_e$
evidence the characteristic lack of information
about the relative phases of the state superposition.
An interesting feature of formula \ref{W_chi_p-m} is that the
pseudo-probability WF can not be concentrated in a too small
region of phase space. We will see that this is equivalent to the
Uncertainty Principle.
%
%
%
\begin{table}
\caption{Left panel: graphical representation of the discrete WF
for one qubit. Right panel: an example of WF for SU(2) coherent
state with $j=1/2$ (up to a factor ${1}/{ 2(1+|\xi|^2
)}$)}\label{tb:Graph_Rep1}
\begin{indented}
\item[]
\medskip
\begin{tabular}{cc}
 \begin{tabular}{|cc|}
  \hline
  $p$ & \begin{tabular}{l|ll}
      1 &  W(0,1) & W(1,1)  \\
      0 &  W(0,0) & W(1,0)  \\\hline
        &    0    &   1     \\
      \end{tabular}\\
  & $q$\\
      \hline
 \end{tabular} &

 \begin{tabular}{|l|l|}\hline
                       $1-Re(\xi)-Im(\xi)$ & $|\xi|^2-Re(\xi)+Im(\xi)$  \\   \hline
                       $1+Re(\xi)+Im(\xi)$ & $|\xi|^2+Re(\xi)-Im(\xi)$  \\   \hline
 \end{tabular}
\end{tabular}
\end{indented}
\end{table}
%
%
%
\begin{table}
\caption{Graphical representation of the discrete WF for two
qubits}\label{tb:Grap_Rep2}
\begin{indented}
\item[]
\medskip
\begin{tabular}{|cc|}
  \hline
  $(p_1,p_2)$ & \begin{tabular}{c|c c c c}
               11 &  W(00,11) & W(01,11) & W(10,11) & W(11,11) \\
               10 &  W(00,10) & W(01,10) & W(10,10) & W(11,10) \\
               01 &  W(00,01) & W(01,01) & W(10,01) & W(11,01) \\
               00 &  W(00,00) & W(01,00) & W(10,00) & W(11,00) \\\hline
                  &    00     &    01    &    10    &    11
                \end{tabular}\\
   & $(q_1,q_2)$\\
  \hline
\end{tabular}
\end{indented}
\end{table}
%
%
\subsection{Axis operators}\label{Prop-axis}
In the single-qubit case the axis operators are defined as
\cite{Wootters}
$\widehat{\xi}_i=\frac{1}{2}\sum_{q,p}\xi_i(q,p) \, \widehat{A}(q,p)$,
where
$\xi_1 (q,p) :=p$, $\xi_2 (q,p):=(q+p)_{mod2}$, and $\xi_3 (q,p) :=q$
while
$\widehat{\xi}_1=\widehat{p}$,
$\widehat{\xi}_2=\widehat{d}$,
and $\widehat{\xi}_3=\widehat{q}$,
(relevant to vertical, diagonal and horizontal lines)
are the vertical, diagonal and horizontal axis operators,
respectively. 
The explicit form of
operators ${\widehat \xi}_i$ reads
\begin{equation}
\label{axis_O_Dbis}
\widehat{\xi}_i=\frac{1}{2}[I-\widehat{S}(i)]\, .
\end{equation}
where $\widehat{S}(i)$ is defined after formula \ref{char_discr_1}.
In this case operators
$\widehat{q}$ and $\widehat{p}$ play the role of (discrete)
\emph{position} and \emph{momentum} operators, respectively.
The spectrum of such axis operators 
is completely determined
by the (two-eigenvalue) spectrum of $\sigma_z$ and
$\sigma_x$, respectively. 
In the sequel operator $\widehat{d}$ will be named
\emph{diagonal-direction} (or simply \emph{diagonal}) operator,
since it is connected to the diagonal lines. Notice that
$\widehat{\xi}_i$ obey commutators
$[\widehat{\xi}_i,\widehat{\xi}_j]=2i\epsilon_{ijk}\widehat{\xi}_k$
showing the SU(2) algebraic structure. Thus they have essentially the
same physical meaning of the three Pauli matrices which is to
describe two-level systems. The only difference is that the relevant
eigenvalues are 0 and 1 (rather than $\pm 1/2$) that are more useful
for treating quantum information applications.
%
%

Analogously to the continuous case, we can define the anticommutator
\begin{equation}\label{symm}
  \{\widehat{\xi}_i,\widehat{\xi}_j\}_S=\frac{1}{2}(\widehat{\xi}_i\widehat{\xi}_j+\widehat{\xi}_j\widehat{\xi}_i)
\end{equation}
(where the label $S$ stands for \textit{standard}). Differently from
the continuous case, the mean value $tr(\rho
\{\widehat{\xi}_i,\widehat{\xi}_j\}_S)$ cannot be written as sum over the
phase-space points of $qpW(q,p)$. We thus introduce an alternative
definition of anticommutator
\begin{equation}\label{symm_d}
  \{\widehat{\xi}_i,\widehat{\xi}_j\}_D=\frac{1}{2}
  (\widehat{\xi}_i+\widehat{\xi}_j-|\epsilon_{ijk}|\widehat{\xi}_k) ,
\end{equation}
where $D$ stands for discrete. This definition allows one to
express the symmetrized product
$\{\widehat{q}_i,\widehat{p}_j\}_D$ as a sum over the phase space
of $W(q,p)$ multiplied by $qp$. In general,
\begin{equation}
\label{mean_v}
  \left< \widehat{\xi}_i \right> =\sum_{q,p}\xi_i W(q,p),  \,\,\,\,
  \left< \{\widehat{\xi}_i,\widehat{\xi}_j\}_D \right>=
\sum_{q,p}\xi_i \xi_j W(q,p)\, .
\end{equation}
The introduction of the symmetrized product
\ref{symm_d} is motivated by the identity $\xi_i
\xi_j=\frac{1}{2}[\xi_i+\xi_j-(\xi_i + \xi_j)_{mod2}]$ (with
$\xi_i =q,p,d \in {0,1}$).

In the two-qubit case ($N=4$) we can perform nine possible
measurements (nine combinations of Pauli matrices),
corresponding to the nine striations of phase space.
This tomographic scheme is not the most efficient since five
orthogonal measurements suffice to determine the state.
Nevertheless, we will consider such scheme (involving
nine \textit{striation} operators
$\widehat{\xi}_i \otimes \widehat{\xi}_j$) in that it
leads to a definition of the WF exhibiting more
interesting entanglement properties.
%
%
\section{Entanglement properties in two qubit systems}\label{Ent}
Given a two-qubit density matrix $\rho$, such a state is said to
be \textit{separable} if there exists a decomposition $\rho =
\sum_k p_k \left|\psi_k\right>\left<\psi_k\right|\otimes
\left|\phi_k\right>\left<\phi_k\right|$ (with the probabilities
$\sum_k p_k =1$). A nonseparable state is said to be entangled. If
a state can be written as a density-matrix product
$\rho=\rho'\otimes \rho''$ ($\rho'$, $\rho''$ relevant to the two
constituent subsystems) then the corresponding WF is written as
$W(q_1,q_2,p_1,p_2)= W_{\rho'}(q_1,p_1)W_{\rho''}(q_2,p_2)$. Thus
the WF associated to a separable state is
\begin{equation}\label{sepw}
  W(q_1,q_2,p_1,p_2)=\sum_{k}p_k W^{'}_k(q_1,p_1)W^{''}_k(q_2,p_2)
\, .
\end{equation}
In this perspective --so far scarcely considered in the
literature-- an entangled state exhibits a WF that cannot be
written in the form \ref{sepw}.
If a classical probability distribution can be written
non-trivially as $p(q_1,q_2,p_1,p_2)=\sum_{i}p_k
p'_k(q_1,p_1)p''_k(q_2,p_2)$, the presence of more than one $p_k
\geq 0$ indicates a (classical) correlation. The WF representation
\ref{sepw} clearly evidences that separable states display a
classical-like correlation since the related WF's have the same
form of a classical distribution, whereas entangled states embody
a different type of correlation named {\it quantum correlation}.
We investigate the entanglement properties of two-qubit WF within
1) the negativity approach 2) a direct reformulation of PT
criterion in terms of WF, 3) the study of non-negativity of WF
relavant both to the density matrix and to its partial trasposed
(deriving from the PT criterion) 3) the Local Uncertainty Relation
(LUR) approach and 4) the Generalized Uncertainty Principle (GUP)
of the continuous case.
%
\subsection{Negativity of WF and entanglement}\label{Ent-neg}
We show that the negativity of $W_{\rho}$ can be connected to the
 non-separability. We give a sufficient condition for
 non-separability, based on the observations of subsection
\ref{Prop-rep}, where it is shown that any single-qubit WF
assumes $(1-\sqrt{3})/8$ as most negative value and $1/2$ as most
positive value.
We can get a two-qubit WF with negative elements considering the
product of WF's of single qubit $W(\alpha_1)W(\alpha_2)$, where
$W(\alpha_1)$ has negative elements, while $W(\alpha_2)$ is
positive. The most negative value of such a two-qubit WF is given
by considering the most negative value for $W(\alpha_1)$ and the
most positive value for $W(\alpha_2)$, as exemplified in table
\ref{tb:Grap_Rep2-examples-neg-separabl}, left panel.
The minimum value we get is $(1-\sqrt{3})/8 \simeq -0.0915$, which is
the lower limit not only for WF relevant to product states. It is easy to
show that any convex combination \ref{sepw} (i.e. separable states) have,
as most negative value, $(1-\sqrt{3})/8$.
However, the value we have found is not in general the most
negative value of a WF, as we can see in table
\ref{tb:Grap_Rep2-examples-neg-separabl}, right panel. The state
represented is the singlet state (a particular the Bell state),
which results to be maximally entangled. Thus if the WF has a
negative value $W(\alpha)< (1-\sqrt{3})/8$, the state is
entangled.
Of course, it a WF has all the values $W(\alpha) \geq
(1-\sqrt{3})/8$, then the state can be entangled or separable. The
Partial Transposition criterion, analyzed in next two sections,
will be useful in such cases.
%
%
\begin{table} \label{tb:most-neg}
\caption{Left panel: graphical representation of the discrete WF
for a two-qubits separable state with the most negative values.
Right panel: graphical representation of the discrete WF for the
singlet state }\label{tb:Grap_Rep2-examples-neg-separabl}
\begin{indented}
\item[]
\medskip
\begin{tabular}{c|c}
\begin{tabular}{ccc}
 \begin{tabular}{|c|c|}\hline
                       $\frac{1}{2}$ & 0  \\   \hline
                       $\frac{1}{2}$ & 0  \\   \hline
       \end{tabular}
       & \begin{tabular}{|c|c|}\hline
                       0.394 & 0.394  \\   \hline
                       -0.183 & 0.394  \\   \hline
       \end{tabular}
       & \begin{tabular}{|c|c|c|c|}\hline
                       0.197 & 0.197 & 0 & 0 \\   \hline
                       -0.0915 & 0.197 & 0 & 0 \\   \hline
                       0.197 & 0.197 & 0 & 0 \\   \hline
                       -0.0915 & 0.197 & 0 & 0 \\   \hline
                     \end{tabular}
                       \\\\
   $W(\alpha_1)$
   & $W(\alpha_2)$
   & $W(\alpha_1)W(\alpha_2)$
\end{tabular}
& \begin{tabular}{c}
  \begin{tabular}{|c|c|c|c|}\hline
                       $-\frac{1}{8}$ & $\frac{1}{8}$ & $\frac{1}{8}$ & $-\frac{1}{8}$  \\   \hline
                       $\frac{1}{8}$  & $\frac{1}{8}$ & $\frac{1}{8}$ & $\frac{1}{8}$ \\   \hline
                       $\frac{1}{8}$  & $\frac{1}{8}$ & $\frac{1}{8}$ & $\frac{1}{8}$ \\   \hline
                       $-\frac{1}{8}$ & $\frac{1}{8}$ & $\frac{1}{8}$ & $-\frac{1}{8}$  \\   \hline
                     \end{tabular}\\\\
                     singlet WF
   \end{tabular}
\end{tabular}
\end{indented}
\end{table}
%
%
%
%
%
\subsection{Partial transposition criterion} \label{Ent-PT}
For all bipartite states (both discrete and continuous), the
well-known partial-transposition (PT) criterion \cite{Peres96,
Horodecki97} turns out to be a necessary condition for separability.
In the $2\times2$ and $2\times3$ dimensional cases, it is also a
sufficient condition.

In the discrete case, the transposition action on a single-qubit
WF and on its characteristic function gives respectively
$W_{\rho^T}(q,p)=W_{\rho}(q,p)-(-1)^{q+p}tr(\widehat{\rho}\widehat{\sigma}_y)$
and $\chi_{\rho^T}(1,1)= -\chi_{\rho}(1,1)$ (where $\chi$ is
unchanged for $(u,v) \ne (1,1)$). In the two-qubit case, the PT
with respect to the second subsystem of $\rho$ provides the new
operator $\rho^{T_2}$ whose matrix elements are $\rho^{T_2}_{m\mu
n\nu}=\rho_{m\nu n\mu}$, where latin (greek) indices refer to the
first (second) subsystem. The WF $W_{\rho^{T_2}}(\alpha_1,
\alpha_2 )$ corresponding to $\rho^{T_2}$ reads
\[\fl
  \frac{1}{4}
  \sum_{m \mu n \nu}\rho_{m\nu n\mu}\, A_{nm}(\alpha_1)\, A_{\mu\nu} ({\alpha_2})
= \frac{1}{4}
  \sum_{m\mu n\nu}\rho_{m\mu
  n\nu} A_{mn} ({\alpha_1}) \, A_{\mu\nu}^{*} (\alpha_2)=
  \frac{1}{4} \, tr \left [ \rho A(\alpha_1) \otimes A^{*}(\alpha_2)\right ]
\, ,
\]
where partial transposition is shown to be equivalent to the
substitution $\widehat{A}_{\alpha_1}\otimes \widehat{A}_{\alpha_2}
\to \widehat{A}_{\alpha_1}\otimes \widehat{A}^{*}_{\alpha_2}$.
Interestingly, the latter is connected to the alternative
definition \cite{Wootters2003,Wootters2004} of WF involving
tomographic properties that differ from those discussed in section
\ref{Prop-axis}. As to the action of PT on a WF and its
characteristic function we find
\begin{equation}\label{PT_discr}
W_{\rho^{T_2}}(\alpha)= W_{\rho}(\alpha)-\tau_{\rho}(\alpha)\, ,
\quad \chi_{\rho^{T_2}}(\alpha_1,11)= - \chi_{\rho}(\alpha_1,11)
\, ,
\end{equation}
where the trace-like term $ \tau_{\rho} (\alpha_1, \alpha_2)
=(-1)^{q_2+p_2} \,
tr(\widehat{\rho}\widehat{A}_{\alpha_1}\otimes\widehat{\sigma}_y)/{2}
$ in equations \ref{PT_discr} embodies the effect of the PT. It is
known that the operator $\rho^{T_2}$ relevant to a separable-state
density operator possesses non-negative eigenvalues.
In view of the properties just discussed,
the PT criterion can be reformulated within the WF approach. If
$\rho^{T_2}$ has all nonnegative eigenvalues, then
$tr(\rho^{T_2}\rho')\geq 0$, for all density matrices $\rho'$,
thus giving
\begin{equation}\label{PPT criterion Wigner}
  \sum_{\alpha}W_{\rho^{T_2}}(\alpha)W_{\rho'}(\alpha)=
  \sum_{\alpha}\chi_{\rho^{T_2}}(\alpha)\chi_{\rho'}(\alpha)\geq 0, \forall
  W_{\rho'}(\alpha),\chi_{\rho'}(\alpha) \, ,
\end{equation}
which is a necessary and sufficient condition for separability. To
illustrate this result, we consider the Werner (mixed) state
$\rho = x | \Psi^{-}\rangle \langle \Psi^{-} | + (1-x) I/4$,
where
$\left|\Psi^{-}\right>=(\left|0,1\right>-\left|1,0\right>)/\sqrt{2}$
and assume $\rho'$ to be the pure state
$\left|\Phi^{+}\right>=(\left|0,0\right>+\left|1,1\right>)/\sqrt{2}$.
It is easy to
show that (see table \ref{tb:Werner_PT})
$\sum_{\alpha}W_{\rho^{T_2}}(\alpha)W_{\Phi^{+}}(\alpha)=(1-3x)/16$,
consistent with the well known separability of the Werner state for
$x\leq {1}/{3}$. This method states that
the separability of $\rho$ is ensured when
inequality \ref{PPT criterion Wigner}
holds for any $W'_{\alpha}$ thus having a limited operational
value. Nevertheless, it is important in that 1) its violation for
some $W'_{\alpha}$ entails that $\rho$ is entangled, and 2) it is
useful to link in a direct way the non-negativity of the WF to the
entanglement properties, as we show in the next section.
\begin{table}
\caption{Application of the PT criterion to a Werner
state.}
\label{tb:Werner_PT}
\begin{indented}
\item[]
\medskip
\begin{tabular}{c|cc}
 $W_{\rho}(\alpha)$ & $W_{\rho^T_2}(\alpha)$ & $W_{\Phi^{+} }(\alpha)$ \\
 \hline\\
\begin{tabular}{|c|c|c|c|}\hline
                       $\frac{1-3x}{16}$ & $\frac{1+x}{16}$  & $\frac{1+x}{16}$  & $\frac{1-3x}{16}$ \\   \hline
                       $\frac{1+x}{16}$ & $\frac{1+x}{16}$ & $\frac{1+x}{16}$ & $\frac{1+x}{16}$  \\   \hline
                       $\frac{1+x}{16}$ & $\frac{1+x}{16}$ & $\frac{1+x}{16}$ & $\frac{1+x}{16}$  \\   \hline
                       $\frac{1-3x}{16}$ & $\frac{1+x}{16}$  & $\frac{1+x}{16}$  & $\frac{1-3x}{16}$ \\   \hline
                     \end{tabular} &
\begin{tabular}{|c|c|c|c|}\hline
                       $\frac{1-x}{16}$ & $\frac{1-x}{16}$ & $\frac{1-x}{16}$ & $\frac{1-x}{16}$  \\   \hline
                       $\frac{1-x}{16}$ & $\frac{1+3x}{16}$ & $\frac{1+3x}{16}$ & $\frac{1-x}{16}$ \\   \hline
                       $\frac{1-x}{16}$ & $\frac{1+3x}{16}$ & $\frac{1+3x}{16}$ & $\frac{1-x}{16}$ \\   \hline
                       $\frac{1-x}{16}$ & $\frac{1-x}{16}$ & $\frac{1-x}{16}$ & $\frac{1-x}{16}$  \\   \hline
                     \end{tabular}
                     & \begin{tabular}{|c|c|c|c|}\hline
                       $\frac{1}{8}$ & $\frac{1}{8}$  & $\frac{1}{8}$  & $\frac{1}{8}$ \\   \hline
                       $\frac{1}{8}$ & $-\frac{1}{8}$ & -$\frac{1}{8}$ & $\frac{1}{8}$ \\   \hline
                       $\frac{1}{8}$ & $-\frac{1}{8}$ & -$\frac{1}{8}$ & $\frac{1}{8}$ \\   \hline
                       $\frac{1}{8}$ & $\frac{1}{8}$  & $\frac{1}{8}$  & $\frac{1}{8}$ \\   \hline
                     \end{tabular}
\end{tabular}
\end{indented}
\end{table}
%
%
%
%
\subsection{Non-negativity of WF and separability}
\label{Ent-PTneg}
We now return to the difficult problem of
establishing a connection between the non-negativity (negativity)
of WF and the separability (non-separability) of the corresponding
state. The starting point consists in observing that any Bell
state has a WF with negative elements, whereas a Werner state has
non-negative WF for any $x\leq 1/3$ (separable cases) (this is
illustrated in table \ref{tb:Werner_PT}). In this respect,
however, we know that exist separable states with negative WF such as
the state in table \ref{tb:most-neg} left panel. On the other hand, one
might conjecture that the non-negativity of WF is a sufficient condition
for separability. Unfortunately, one can show that non-negative WF's
exist which correspond to entangled states. As a possible strategy
for solving this problem, we thus propose a simple method
based on considering the non-negativity features of the WF's relevant both
to $\rho$ and to $\rho^{T_2}$ to check the separability of the
state.

Let us assume that, given a state $\rho$ with $W_{\rho}$ and
$W_{\rho^{T_2}}$ non-negative for every phase-space point
$\alpha$, there is a $W_{\rho'}$ giving
$\sum_{\alpha}W_{\rho^{T_2}}(\alpha)W_{\rho'}(\alpha)<0$ (this
entails, using \ref{PPT criterion Wigner}, that the state is
entangled). We show that these assumptions lead to a
contradiction. First we note that the last inequality can be
rewritten as
$ \sum_{\alpha}W_{\rho}(\alpha)W_{\rho'}(\alpha)<
\sum_{\alpha}\tau_{\rho}(\alpha)W_{\rho'}(\alpha)$.
On the other hand, the non-negativity of both $W_{\rho}$ and
$W_{\rho^{T_2}}$, and  equation \ref{PT_discr} [which gives
$W_{\rho}(\alpha)\geq \tau_{\rho}(\alpha)$ for all $\alpha$] imply
that $\sum_{\alpha}W_{\rho}(\alpha)W_{\rho'}(\alpha) \geq
\sum_{\alpha}\tau_{\rho}(\alpha)W_{\rho'}(\alpha)$, which clearly
involves a contradiction.

It follows that, given a state $\rho$, if both $W_{\rho}$
\textit{and} $W_{\rho^{T_2}}$ have non-negative elements, than
$\rho$ is separable. Viceversa, if a state $\rho$ is entangled,
then $W_{\rho}$ \textit{or} $W_{\rho^{T_2}}$ has negative values.
Such a result --which is a necessary condition to ensure
entanglement (sufficient condition for separability)-- relates the
nonclassic character of entangled states to the presence of
negative elements in $W_{\rho}$ and $W_{\rho^{T_2}}$. This
criterion has been confirmed by testing it on thousands of
randomly-generated density matrices and on the Werner state (for
$x \leq 1/3$ both $W_{\rho}$ and $W_{\rho^T_{2}}$ are positive,
which implies separability).
%
\subsection{Local Uncertainty Relation}
\label{Ent-LUR}
%
%
\begin{table}
\caption{Two examples of covariance matrix of two qubit WF}
\label{tb:tabella}
\begin{indented}
\item[]
\medskip
\begin{tabular}{c|c}
 Werner State & $\left|\Phi^{+}\right>$ \\ \hline \\
                     $\left[\begin{tabular}{cccccc}
                       $\frac{1}{4}$ & 0 & 0 & $-\frac{x}{4}$ & 0 & 0 \\
                       0 & $\frac{1}{4}$ & 0 & 0 & $-\frac{x}{4}$ & 0 \\
                       0 & 0 & $\frac{1}{4}$ & 0 & 0 & $-\frac{x}{4}$ \\
                       $-\frac{x}{4}$ & 0 & 0 & $\frac{1}{4}$ & 0 & 0 \\
                       0 & $-\frac{x}{4}$ & 0 & 0 & $\frac{1}{4}$ & 0 \\
                       0 & 0 & $-\frac{x}{4}$ & 0 & 0 & $\frac{1}{4}$ \\
                     \end{tabular}\right]$
                     & $\left[\begin{tabular}{cccccc}
                       $\frac{1}{4}$ & 0 & 0 & $\frac{1}{4}$ & 0 & 0 \\
                       0 & $\frac{1}{4}$ & 0 & 0 & $-\frac{1}{4}$ & 0 \\
                       0 & 0 & $\frac{1}{4}$ & 0 & 0 & $\frac{1}{4}$ \\
                       $\frac{1}{4}$ & 0 & 0 & $\frac{1}{4}$ & 0 & 0 \\
                       0 & $-\frac{1}{4}$ & 0 & 0 & $\frac{1}{4}$ & 0 \\
                       0 & 0 & $\frac{1}{4}$ & 0 & 0 & $\frac{1}{4}$ \\
                     \end{tabular}\right]$ \\\\
\end{tabular}
\end{indented}
\end{table}
%
%
%

%
It is known that the violation of local uncertainty relations
(LUR's) is a signature of entanglement \cite{HofTak, Guhne}. Given
two qubits 1 and 2, the inequalities
\begin{equation}\label{LUR1}
  \sum_{i} U[\widehat{\xi}_{i}^{(1)}] \geq \frac{1}{2} \,\,\,\,
  \textrm{and}\,\,\,
  \sum_{i} U[\widehat{\xi}_{i}^{(2)}] \geq \frac{1}{2} \,   ,
\end{equation}
are known to be uncertainty relations relevant to the single qubit
systems $k=1,2$, where $U[\widehat{\xi}_{i}^{(k)}]
=\left<(\widehat{\xi}^{(k)}_{i})^2\right>-\left<\widehat{\xi}_{i}^{(k)}\right>^2$
are the uncertainties relevant to the set of axis operators
${\widehat{\xi}_i}^{(k)}$ defined in formula \ref{axis_O_Dbis}. We
note that simple calculations prove the equivalence between
formula \ref{LUR1} and \ref{W_chi_p-m}, entailing that the
pseudo-probability can not be concentrated in a too small region
of the phase space.
As shown in \cite{HofTak}, in the two-qubit case, separable states
are constrained by the single-qubits uncertainty relations
\begin{equation}\label{LUR2}
\sum_{i} U[\widehat{\xi}_{i}^{(1)}\otimes I + I\otimes
\widehat{\xi}_{i}^{(2)}]\geq 1 \, .
\end{equation}
As a consequence, a state appears to be entangled if inequality
\ref{LUR2} is violated.

LUR inequalities can be formulated in terms of WF, by defining the
first-order covariance matrix of single-qubit WF
\begin{equation}\label{Covariance_1}
 V_{i
 j}^{(X)}=\left<\{\Delta\widehat{\xi}_i, \Delta\widehat{\xi}_j\}_X\right>=\left<\{\widehat{\xi}_i\widehat{\xi}_j\}_X\right>-\left<\widehat{\xi}_i\right>\left<\widehat{\xi}_j\right> ,
\end{equation}
where we use the axis operators
${\widehat{\xi}_i}={\widehat{p},\widehat{d},\widehat{q}}$, and
$\Delta\widehat{\xi}_i =
\widehat{\xi}_i-\left<\widehat{\xi}_i\right>$.
Moreover, the label $X=S,D$, linking to the two types of
anticommutator defined in formulas \ref{symm} and \ref{symm_d},
leads to two different covariance matrices. Nevertheless, the
following results are independent from definition of anticommutator,
and we will write the parameter $X$ only when necessary. Recalling
that matrix $V_{ij}$ is a $3\times3$ semi-definite positive symmetric
matrix and that diagonal elements $V_{ii}$, named \textit{variances},
coincide with the uncertainties $U[\xi_i]$, then the sum of
diagonal elements $V_{ii}$ is positive, consistent with
\ref{LUR1}.

Following the scheme of reference \cite{Simon_99} for the continuous
case, the covariance matrix of two qubits is built by writing formula
\ref{Covariance_1} with the enlarged set $\widehat{\xi}=
(\widehat{p}_1,\widehat{d}_1,\widehat{q}_1,\widehat{p}_2,\widehat{d}_2,
\widehat{q}_2)$ giving a $6\times 6$ matrix. A compact version of
covariance matrix is given by
\begin{equation}\label{Variance_matrix}
V= \left [
\begin{tabular}{cc}
    A & C \\
    $C^T$ & B
  \end{tabular}
\right ]\, ,
\end{equation}
where $A$, $B$, $C$ are $3\times 3$ matrices. Notice that matrix
elements of $A$ and $B$ represent the covariance matrix $V_{ij}$
relevant to qubit 1 and to qubit 2, respectively, while matrix $C$
represents the inter-qubit correlations between axis operators
$\widehat{\xi}_{i}^{(1)}$ and $\widehat{\xi}_{i}^{(2)}$. Two-qubit
covariance matrix can be easily computed by means of equations
\ref{mean_v} once the WF is known. It is easy to show that the
inter-qubit correlations $C_{ii}$ measure the degree of correlation
between spin observables $\widehat{\sigma}_i^{(1)}$ and
$\widehat{\sigma}_i^{(2)}$. Hence their
operational meaning is the establish the interdependence of the
two constituent subsystems.
At this point, the WF formulation of LUR's is easily achieved.
Upon observing that
\[ \fl
  U[\widehat{\xi}_{i}^{(1)}\otimes I + I\otimes
  \widehat{\xi}_{i}^{(2)}]=U[\widehat{\xi}_{i}^{(1)}]+
  U[\widehat{\xi}_{i}^{(2)}]+
  2 \{\left<\widehat{\xi}_{i}^{(1)}\otimes
  \widehat{\xi}_{i}^{(2)}\right>-
  \left<\widehat{\xi}_{i}^{(1)}\right>\left<\widehat{\xi}_{i}^{(2)}\right>\} \, ,
\]
the LUR relevant to the axis operators becomes
\begin{equation}\label{LUR2_variance}
  trA + trB + 2tr C \geq 1 \, ,
\end{equation}
where only diagonal elements of submatrices are involved, thus
making the formula independent from the definition of
anticommutator.
This equation has the following interpretation: if the
correlations $C_{ii}$ are negative and their absolute values are
sufficiently large, than the inequality is violated and the state
is entangled. This evidences that non-separability strongly
depends on the inter-qubit correlations described by $tr \, C$. An
important problem that deserves to be clarified is raised by those
entangled states where correlations $C_{ii}$ are positive. To
answer to this question, in table \ref{tb:tabella} we consider
covariance matrices relevant to both the WF of Werner's state
(including as well the singlet state with weight $x$) and the WF
of the Bell state
$\left|\Phi^{+}\right>=(\left|00\right>+\left|11\right>)/\sqrt{2}$.
In the first case, formula \ref{LUR2_variance} is violated for
$x \leq 1/3$, which is a correct result. Instead, in the second
case, formula \ref{LUR2_variance} is not violated, thought the
Bell state is known to be maximally entangled. We can show that,
except for the singlet state, no Bell state violates formula
\ref{LUR2_variance} and the criterion does not supply
information about separability. The singlet case (corresponding in
the table \ref{tb:tabella}
to the $x=1$ case)
differs from the other Bell states in that all the diagonal
elements of matrix $C$ are negative. The other Bell states have
elements $C_{ii}$ with alternating sign, which makes the violation
of formula \ref{LUR2_variance} impossible. Nevertheless, it is
clear that such correlations, thought not negative, contain a
large amount of information on non-separability. We thus propose
the following modified inequality as a necessary condition for
separability
\begin{equation}
\label{LUR2_variance_gen}
  \sum_i |C_{i,i}| \leq \frac{trA + trB -1}{2} \, ,
\end{equation}
whose main feature is to replace the diagonal elements of $C$ with
their absolute values. The effectiveness of formula
\ref{LUR2_variance_gen} is confirmed by the fact that it is
violated by any Bell states. The quantum-mechanical meaning is
also clear in that, if a state is separable, then the absolute value of
the correlations must be bounded from above.
%
We easily prove that this inequality follows from equation
\ref{LUR2} by resorting to the more general operator
$\widehat{\xi}_{i}^{(1)}\otimes I + \epsilon_i I\otimes
\widehat{\xi}_{i}^{(2)}$ where $\epsilon_i = \pm 1$. With such a
choice, the LUR condition written in terms of covariance matrix
reads $tr A+trB+2\sum_i \epsilon_i C_{i,i} \geq 1$. To obtain
inequality \ref{LUR2_variance_gen}, it is sufficient to consider
$\epsilon_i=-1$ when $C_{ii}$ is positive. As a final comment, we
notice that the sum of correlations \ref{LUR2_variance_gen} thus
exhibits an upper limit for separable states:
entangled states may overcome it, and the exceeding part is an
indicator of \textit{quantum correlation}.
\subsection{Generalized Uncertainty Principle (GUP) and PT criterion}
\label{Ent-GUP}

In the continuous case, a well-known separability criterion
\cite{Simon_97,Simon_99} is obtained by combining the Generalized
Uncertainty Principle (GUP) with the application of the PT
criterion to the variance matrix relevant to position and momentum
operators. For a system of two (one-dimensional) particles in a
continuous space, the GUP-based criterion states that, if a state
$\rho$ is separable, one can construct a matrix $M= tr(\rho \xi_i
\xi_j)$ which is semi-definite positive under PT, namely
\cite{Simon_99}
\begin{equation}\label{uncertainty_continuous}
  M=V+\frac{i}{2}\Omega \geq 0 \, ,
\end{equation}
where
$V_{\alpha\beta}=\left<\{\Delta\widehat{\xi}_{\alpha},\Delta\widehat{\xi}_{\beta}\}\right>$
is the covariance matrix,
$\Delta\widehat{\xi}_{\alpha}
=\widehat{\xi}_{\alpha}-\left<\widehat{\xi}_{\alpha
}\right>$,
$\widehat{\xi}_\alpha = \{ \widehat{q}_1 , \, \widehat{p}_1 \}$
and
$[\widehat{\xi}_{\alpha},\widehat{\xi}_{\beta}]=i\Omega_{\alpha\beta}$
with
$$
  \Omega = \left [
  \begin{tabular}{cc}
    J & 0 \\
    0 & J
    \end{tabular} \right ] \, , \quad
    J= \left [ \begin{tabular}{cc}
    0 & 1 \\
    -1 & 0
    \end{tabular} \right ] \, .
$$
Following the PT criterion, given a separable state
$\widehat{\rho}$ and its WF $W_{\rho}$, the PT generates a
nonnegative operator ${\widehat \rho}^{T_2}$ and a genuine WF
$W_{\rho^{T_2}}$ still satisfying the equation \ref{uncertainty_continuous}.

The extension of the previous GUP-based criterion to the discrete
case requires that each separable state can be associated to a
matrix $M$ semi-definite positive under PT.
Considering first the single-qubit case, we define the matrix
$M_{ij}=[tr(\widehat{\rho} \widehat{\xi}_i \widehat{\xi}_j )]$
written in terms of the list of operators
${\widehat{\xi}_i}={\widehat{I},\widehat{p},\widehat{d},\widehat{q}}$.
It can be easily shown that $\rho\geq 0$ entails $M\geq 0$. The
latter is equivalent to the condition
\begin{equation}
\label{GUP_discrete1} V_{jk}^{S}+\frac{i}{2}\epsilon_{jkl}\chi_l
\geq 0 \, ,
\end{equation}
where covariance matrix $V^{(S)}$ in equation \ref{Covariance_1},
in the present case, is a $3\times3$ matrix and is related to the
standard definition of anticommutator \ref{symm}.
Condition \ref{GUP_discrete1} implies $tr(V)\geq {1}/{2}$, which
is equivalent to the LUR equation for  single qubit \cite{HofTak}.
It is worth observing how any other choice for the set $\{\xi_i\}$
implies that $M\geq 0$ iff $\rho \geq 0$ provided $\{\xi_i\}$
forms a complete basis of the space of hermitian matrices for a
single-qubit.

In the case of two qubits, once more in analogy with the
continuous case, it seems quite natural to derive $M$ from the set
$\widehat{\xi}=(I,\widehat{p_1},\widehat{d_1},\widehat{q_1},\widehat{p_2},\widehat{d_2},\widehat{q_2})$.
Following the standard prescriptions \cite{Simon_99} for
calculating the GUP inequality, we have that $M\geq 0$ and,
equivalently,
\begin{equation}\label{GUP_discrete2}
\left [
\begin{tabular}{cc}
$A_{jk}+\frac{i}{2}\epsilon_{jkl}\xi^{(1)}_{l}$ & $C_{jn}$\\
$C_{km}$ & $B_{mn}+\frac{i}{2}\epsilon_{mns}\chi^{(2)}_{s}$
\end{tabular} \right ] \geq 0\, .
\end{equation}
The matrix on the left-hand side is a
$6\times 6$ matrix that can be written in terms of the $3\times 3$
matrices $A$, $B$, $C$ appearing in equation \ref{Variance_matrix}.
Similar calculations show how $\widetilde{M} = tr (\rho^{T_2} \, \xi_i \xi_j)$
is such that $\widetilde{M} \geq 0$ if $\rho^{T_2} \geq 0$.
Then we conclude that the separability condition for a state
$\rho$ (achieved within the PT criterion when both $\rho \geq 0$
and $\rho^{T_2} \geq 0$ are satisfied) is now ensured by ${M} \geq
0$ and $\widetilde{M} \geq 0$. Notice that condition
$\widetilde{M} \geq 0$ can be reduced as well to the equivalent
form
\begin{equation}\label{GUP_discrete2_PT}
 \left [
\begin{tabular}{cc}
$A_{jk}+\frac{i}{2}\epsilon_{jkl}\chi^{(1)}_{l}$ & $\widetilde{C}_{jn}$\\
$\widetilde{C}_{km}$ &
$\widetilde{B}_{mn}+\frac{i}{2}\epsilon_{mns}\widetilde{\chi}^{(2)}_{s}$
\end{tabular} \right ]
\geq 0\, ,
\end{equation}
where
$\widetilde{B}$, $\widetilde{C}$ are
determined using once more the PT operation.
Formula \ref{GUP_discrete2_PT} containing the axis-operator
covariance matrix is the core of the two-qubit PT criterion. In
table \ref{tb:GUP_example}, we illustrate the application of the
present criterion to the Werner state. In this case $V=M$ and the
eigenvalues of matrix $\widetilde{M}$ (relevant to $\rho^{T_2}$)
are positive for $x \geq 1$ (rather than for $x \geq 1/3)$.
Unfortunately, this means that the GUP is not violated so that the
criterion does not give information about separability. This can
be explained with the fact that, when using the set of operators
$(\widehat{p_1},\widehat{d_1},\widehat{q_1},\widehat{p_2},\widehat{d_2},\widehat{q_2})$,
the nonnegativity of $\widetilde{M}$ 
is only a necessary condition for separability. In order to cure
this problem we have generalized the GUP-based criterion by using
the enlarged set of operators $\widehat{\xi}_i\otimes
\widehat{\xi}_j$, which leads to a $9\times 9$ matrix $M$. In this
case, we could have a violation of the positivity condition under
partial transposition. This result will be discussed in a separate
paper.
\begin{table}
\caption{An example of GUP in the two-qubit case.}\label{tb:GUP_example}
\begin{indented}
\item[]
\medskip
\begin{tabular}{cc}
 $M$ & $\widetilde{M}$ \\
                     $\left[\begin{tabular}{cccccc}
                       $\frac{1}{4}$ & 0 & 0 & $-\frac{x}{4}$ & 0 & 0 \\
                       0 & $\frac{1}{4}$ & 0 & 0 & $-\frac{x}{4}$ & 0 \\
                       0 & 0 & $\frac{1}{4}$ & 0 & 0 & $-\frac{x}{4}$ \\
                       $-\frac{x}{4}$ & 0 & 0 & $\frac{1}{4}$ & 0 & 0 \\
                       0 & $-\frac{x}{4}$ & 0 & 0 & $\frac{1}{4}$ & 0 \\
                       0 & 0 & $-\frac{x}{4}$ & 0 & 0 & $\frac{1}{4}$ \\
                     \end{tabular}\right]$
                     & $\left[\begin{tabular}{cccccc}
                       $\frac{1}{4}$ & 0 & 0 & $-\frac{x}{4}$ & 0 & 0 \\
                       0 & $\frac{1}{4}$ & 0 & 0 & $\frac{x}{4}$ & 0 \\
                       0 & 0 & $\frac{1}{4}$ & 0 & 0 & $-\frac{x}{4}$ \\
                       $-\frac{x}{4}$ & 0 & 0 & $\frac{1}{4}$ & 0 & 0 \\
                       0 & $\frac{x}{4}$ & 0 & 0 & $\frac{1}{4}$ & 0 \\
                       0 & 0 & $-\frac{x}{4}$ & 0 & 0 & $\frac{1}{4}$ \\
                     \end{tabular}\right]$ \\\\
\end{tabular}
\end{indented}
\end{table}
%
%
\section{Conclusions}
In the present work, we have considered the WF defined in
\cite{Wootters} focusing our attention on two properties of the
two-qubit WF, the negativity and the covariance matrix, which are
useful in the characterization of entanglement. After
reformulating/generalizing the PT, LUR, and GUP-based separability
criteria in the WF formalism, we have tried to evidence what
features of the WF and of its covariance matrix are able to reveal
the presence of entanglement.

In section \ref{Ent-neg} we have found that a two-qubit WF
relevant to a separable state can not assume values lower than
$(1-\sqrt{8})/4$. In section \ref{Ent-PT}, we have recast the PT
criterion \ref{PT_discr} in terms of WF by means of inner-product 
rule \ref{orth_wign_discr}. Based on this result, in
section \ref{Ent-PTneg} we have shown that the non-separability of
$\rho$ entails the presence of negative elements in $W_{\rho}$ or
in $W_{\rho^{T_2}}$. Interestingly, these facts relate the main
non-classical feature of the WF (the presence of negative values)
to the presence of entanglement in the two-qubit system.
Considering the separability problem within the LUR criterion, in
section \ref{Ent-LUR} we have reformulated it in terms of
covariance-matrix elements of WF \ref{LUR2_variance}. In
particular, we have found a stronger version of the LUR criterion
(illustrated by formula \ref{LUR2_variance_gen}) once more
involving the covariance matrix. This generalized criterion, which
has been tested both on Bell states and on Werner states,
evidences that the presence of strong correlations can be used to
detect non-separability.
Finally, in section \ref{Ent-GUP}, we have studied the
analogue of the GUP-based separability criterion
(continuous case) from the viewpoint of discrete WF's.
We have shown that adopting the same procedure of the
continuous case leads to criterion \ref{GUP_discrete2_PT}.
The latter does not succeed in detecting entanglement
as a consequence of the fact that the set of operators used
to build the discrete GUP is too small.
In order to cure this problem,
we have enlarged such an operator set thus obtaining that
${\tilde M} \geq 0 \Leftrightarrow \rho^{T_2} \geq 0$.
Such an equivalence provides the basis to extend in an
effective way the GUP-based separability criterion
from the continuous to the discrete case.

Future work about entanglement properties of the two-qubit WF will
be developed in two directions. Our first objective is to derive,
relying on equation \ref{GUP_discrete2_PT}, the explicit form of
a generalized GUP-based separability criterion from a
suitably enlarged operator set.
A second important problem which deserves to be deepen
is to establish how the presence of negative
elements in WF's $W_{\rho}$ (and $W_{\rho^{T_2}}$) relevant to
entangled states is related to the violation of inequality
\ref{LUR2_variance_gen} issued from the LUR condition.
Such aspects will be investigated in a separate paper.
%
%

\section*{References}

%
\end{document}